# AUTOREGULATION OF THE TOTAL NUMBER OF CELLS IN CULTURE

A. A. Kozlov

*Biological Faculty, Tbilisi State University, Tbilisi, 380043 Georgia*

*E-mail: alekskoz@rambler.ru*

## Initial hypothesis

According to the universally accepted concept of the development of life on the Earth, multicellular organisms initially emerged as a result of either the union of identical unicellular organisms with the following functional differentiation, or the union of symbionts, in which there already was a certain simple functional separation. However, in either case the progenitors of multicellular organisms were ensembles, communities of unicellular organisms. For a certain number of unicellular organisms to be treated as an ensemble, there must be some interconnection between its members. Colonies of mechanically connected unicellular organisms were a later, more advanced stage; here, unicellular organisms living separately are considered. Such interconnection must, in particular, limit from above the total numbers of the members of the ensemble, because an excessive increase in these numbers could disturb the connections between members of the ensemble to the extent of its destruction. In addition, too large numbers of members in the ensemble could lead to nutrient depletion in its habitat.

One can assume that such interconnection between unicellular organisms was evolutionarily developed and genetically fixed.

I assumed that modern unicellular organisms retain such an ability to regulate their total numbers. The validity of this assumption was tested in experiments, whose results are presented in this paper.

**Object and method**

Objects of investigation were infusoria *Colpoda sp.* and, in some experiments, *Paramecia sp*. Infusorial cultures were grown in a hay decoction (2 g of dry sweet clover *Melilotus officinalis* per liter of ordinary water) at room temperature (22-24°C) and ambient illumination in cylindrical glass vessels of different sizes.

In each experiment, once every day at the same time (12 noon) the average cell concentrations in all the experimental vessels were measured and plotted against time. The nutrient medium was neither replaced nor refreshed.

**Results**

In the first set of experiments, infusoria *Colpoda* sp. cultures were grown in three vessels of volume $V_1 = 30$ cm$^3$ each and three vessels of volume $V_2 = 3$ cm$^3$ each. A total of five experiments were performed. Figure 1 presents the results of the second experiment of this test run.

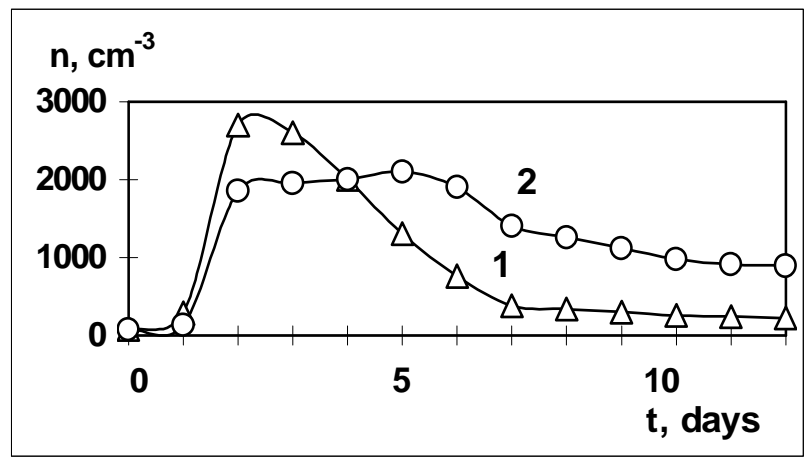

Fig. 1. Cell concentrations n in the volumes $V_1$ (1) and $V_2$ (2) in the second experiment

Let us note three important features of the curves in Fig. 1, which will be discussed below.

(1) The cell concentration in the volume $V_1$ increases and attains a maximum, whereupon it rapidly decrease by $(30.0\pm4.1)$% per day. Later, the decrease in the concentration abruptly slows down and then is (8.6±1.2)% per day. The behavior of the cell concentration in the volume $V_2$ is similar but much less pronounced.

(2) In the phase of the slow decrease in the cell density (let us call it the "quasi-stationary" phase), the ratio between the cell concentrations in the volumes $V_2$ and $V_1$ is almost constant and is $3.9 \pm 0.3$.

(3) In the first several days of the cell development, the cell concentration in the larger volumes rises more rapidly than that in the smaller volumes.

The results of the other four experiments were qualitatively identical.

The second experiment had a continuation. On the 12th day of cultivation, the cell suspensions in the three vessels of volume $V_1$ were thoroughly stirred, and the suspension from one of these vessels was used to fill in three vessels of volume $V_2$. (Moments of transfusion are marked by arrows on Fig. 2)

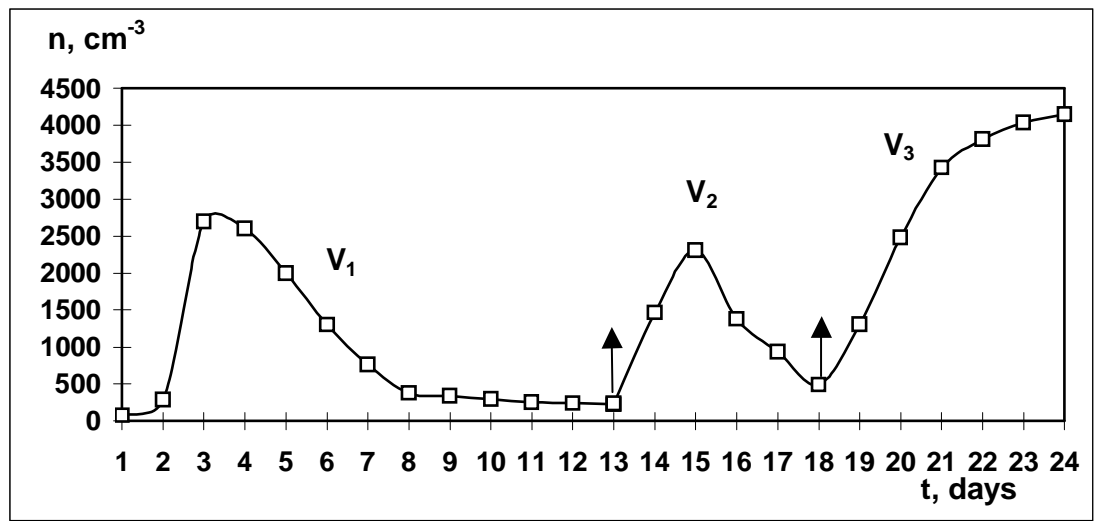

Fig. 2 Changes in cell concentrations n after their transfusion from vessel $V_1$ to vessels $V_2$, and then from vessel $V_2$ to vessel $V_3$ as function of time t. Moments of transfusion are marked by arrows

As one can see from Fig. 2, the cell concentrations in the new vessels of volume $V_2$ began to rapidly increase at once, and on the eighth day, the increase in the concentration ended with the quasi-stationary phase.

A similar procedure was next performed with a culture grown in three vessels of volume $V_2$. By that time, the infusoria *Colpoda* sp. concentration was $(488\pm52)$cm$^{-3}$. The cell suspensions in the three vessels were thoroughly stirred, and the contents were transfused into three vessels of volume $V_3 = 0.9$ cm$^3$.

As one can see from Fig. 2, like after the first transfusion, cell concentrations in vessels of volume $V_3$ began to rapidly increase.

However, a further (after the fourth day) increase was most likely to be due to the decrease in the cell suspension volume (0.9 cm$^3$) because of sampling for finding the cell concentration. Three samples of a volume of 0.02 cm$^3$ each per one experimental point were taken, because of which the cell suspension volume by the end of the experiment decreased down to 0.54 cm$^3$.

The results of these experiments suggested the following.

(4) A decrease in the cultivation volume accompanied by a proportional decrease in the numbers of cells in this volume causes a wave of cell division.

In the next experiment, infusoria *Colpoda* sp. were cultivated in ten vessels of volume $V_2 = 3$ cm$^3$ each. Once the cells reached the quasi-stationary phase, the contents of these vessels were transfused into a single vessel of volume $V_1 = 30$ cm$^3$. Figure 3 illustrates the change in the cell concentration as a function of time elapsed from the transfer of the cell suspension.

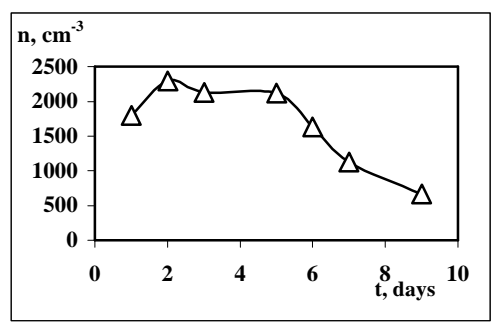

Fig. 3 Cell concentration n as a function of time t elapsed from the transfer of the cell suspension (see text)

Note that, after an insignificant increase on the fourth day, the cell concentration started to drop rapidly (by 27% per day).

This suggests the following.

(5) An increase in the cultivation volume accompanied by a proportional increase in the numbers of cells in this volume abruptly decreases the cell concentration.

The results of these experiments are indicative of the fact that, in a culture of unicellular organisms, there is a certain mechanism that regulates the cell concentration, depending on the cultivation volume.

One can assume that this mechanism tries to keep the constancy of the total numbers of the cells in this culture even after a change in the cell suspension volume. The validity of this assumption was checked in the following experiment.

Infusoria *Colpoda* sp. were cultivated in three vessels of volume $V_9 = 6$ cm$^3$ each. On the 40th day of cultivation, the suspensions of cells in the quasi-stationary phase of their development (the average numbers of the cells in each of the three vessels were $3500 \pm 468$) were thoroughly stirred, whereupon much of the cell suspension was removed to leave only 0.9 cm$^3$ of the suspension in each vessel. At that moment, in each of the three vessels, the average cell concentration was $(583\pm78)$cm$^{-3}$, and the average numbers of the cells were $525\pm70$. Figure 4 shows a further change in the total numbers of the cells in these vessels.

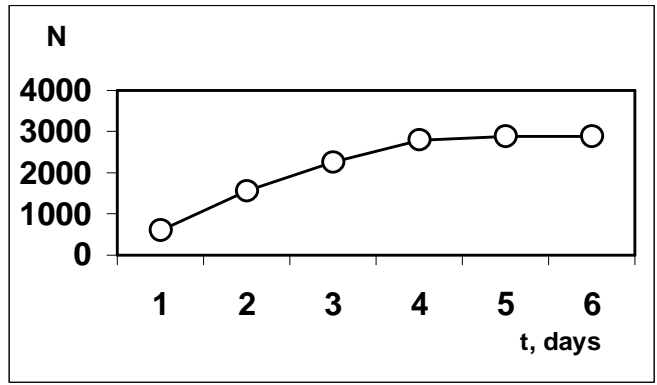

Fig. 4 Total numbers N of cells in the volumes 0.9 cm$^3$ as a function of time t elapsed from the removal of much of the cell suspension (see text)

As one can see from Fig. 4, the total numbers of the cells in the suspensions of the volume 0.9 cm$^3$ each reaches 86% of the initial total numbers of the cells in the suspensions of the volume 6 cm$^3$ each. This resembles the regenerative growth of an organ after its partial resection in multicellular organisms.

Thus, the results of this experiment confirmed the assumption that, in a culture of unicellular organisms, there is a mechanism regulating the total numbers of cells so that these numbers can be maintained approximately constant even if the cultivation volume is changed (at least within a certain range of volumes).

Let us go back to Fig. 1. A decrease in the cell concentration after the culture reaches its maximal development is generally accepted to be because of nutrient depletion in its habitat and habitat pollution with metabolic products. This explanation was tested in an experiment on a stock infusoria *Colpoda* sp. culture, which was maintained in the exponential growth phase by daily dilution with a fresh nutrient medium. Beginning with the phase of the rapid decrease in the cell concentration in the test culture in volume $V_1$, a suspension sample of about 0.3 cm$^3$ in volume was taken (to weaken the effect of sampling, several cell suspensions of volume $V_1$ were used). From the sample, all infusoria were removed with a micropipette (all other means of removing cells from a medium, e.g., filtration, centrifugation, heating, etc., could un-predictably change the composition of the nutrient medium). Into this, cell-free, already used nutrient medium, infusoria from the stock culture were transferred. Such a procedure was performed once every three days until that time, when the culture in volume $V_1$ was already in the quasi-stationary phase. In a reference experiment, infusoria were transferred into a fresh nutrient medium. In either used or fresh medium, infusoria developed virtually identically.

**On the possible physical channel of connection between cells**

Since a substantial part of the experimental data obtained cannot be explained by any chemical mechanism, I assumed that there is a physical coupling channel between cells, which runs or terminates one regulatory program or another. In particular, along this coupling channel, each cell must receive information on the total numbers of cells in the cultivation volume.

It is known that any cell emits electromagnetic radiation in a certain spectral region [1, 2]. There is also evidence that cells emit acoustic radiation [3, 4]. These types of radiation can, in principle, carry information on the total numbers of cells in the cultivation volume. I assumed that, when the radiation flux emitted by all the cells in the cultivation volume attains a certain critical value, the cell division ceases. Let us consider this assumption in greater detail.

Let in a vessel with a liquid nutrient medium there is a homogeneous suspension of cells. Let each cell is an isotropic source of radiation (electromagnetic or sound). Then the average density of radiation in volume will be proportional to number of sources (i.e. - to volume) and inversely proportional to its square of a surface (due to losses of radiation through a surface). If the form of cylindrical volume slightly differs from sphere ($2R \approx H$), the average density of radiation inside volume will be equal to:

$$\psi = knV/S, \qquad\qquad 1)$$

where k - factor of proportionality determining radiating ability of each cell, n - concentration of cells in volume, V and S - volume and the square of its surface accordingly. Having substituted values V and S for cylindrical volume, we will obtain:

$$\psi = knRH/2(R+H), \qquad 2)$$

where R and H - radius and height of volume accordingly.

If our assumption is fair, critical density of radiation at which cell division stops, in the big volumes will be achieved at smaller concentration of cells, than in small. We shall take two cylindrical volumes with different sizes $V_1$ and $v_2$ with radiuses R and r and heights H and h. (Bigger letters concern to the greater volume sizes).

Then $\psi_1 = kn_1RH/2(R+H)$ and $\psi_2 = kn_2rh/2(r+h)$ - average density of radiation inside these volumes. As it is supposed, that cell division stops in these different volumes at identical average density of radiation in them, from a condition $\psi_1 = \psi_2$ it is possible to find the attitude the of concentration of cells in these volumes in quasi-stationary phase:

$$n_1/n_2 = RH(r+h)/rh(R+H) \qquad 3)$$

The table compares the ratios between the concentrations of the cells in the quasi-stationary phase in different volumes as calculated by expression 3) and found experimentally.

Table

Comparison of the concentrations of the cells in quasi-stationary phase for the different volumes calculated by the formula 3) with the concentrations defined experimentally.

| Volume (Object) | R (cm) | H (cm) | V (cm$^3$) | Comparison (S/L) | $n_S/n_L$ (calc.) | $n_S/n_L$ (experiment) |
|---|---|---|---|---|---|---|
| $V_1$(Colpoda) | 1.6 | 3.7 | 30.0 | 2 - 1 | 2.5 | 3.9 ± 0.3 |
| $V_2$(Colpoda) | 0.5 | 3.8 | 3.0 | | | |
| $V_4$(Colpoda) | 0.35 | 2.0 | 0.8 | 4 - 5 | 3.2 | 3.1 ± 0.2 |
| $V_5$(Colpoda) | 1.25 | 4.0 | 19.6 | | | |
| $V_6$(Colpoda) | 1.75 | 2.0 | 19.2 | 4 - 6 | 3.1 | 3.2 ± 0.7 |
| $V_1$(Paramecia) | 1.6 | 3.7 | 30.0 | 1 - 7 | 1.5 | 1.6 ± 0.5 |
| $V_7$(Paramecia) | 2.5 | 5.0 | 100.0 | 8 - 1 | 2.2 | 2.5 ± 0.8 |
| $V_8$(Paramecia) | 0.8 | 1.5 | 3.0 | 8 - 7 | 3.2 | 4.0 ± 1.1 |

The note to the table: in columns 5, 6 and 7 indexes S and L concern to smaller and large volumes accordingly.

**Proposed explanations of experimental data**

The proposed physical regulation mechanism is most likely to affect the cell membrane complex. This complex can be regarded as a cooperative system with two stable conformational states and one unstable conformational state. At the first stable state, the cell arrives immediately after the cell division is completed. A transition to the second stable state requires a starting energy of about 5 eV [1, 6]. This energy triggers the generalized conformational rearrangement of the cell membrane complex, owing to which the cell becomes capable of advancing further along the cell division cycle. When the cell is in the first stable state, the radiation regulating the numbers of cells can bring it into the third state, in which the cell leaves the cycle. However, this state is so unstable that even local temperature fluctuations return the cell to the first stable state. Yet, when the radiation flux density attains its critical value as the overall number of the cells in the culture increases, the cell has no time to return to the stable state. However, vigorous shaking can probably take a portion of the cells out of this unstable state for a while.

The considered mechanism of regulating the numbers of cells in a culture (colony) is necessary for heterotrophs, because an excessive increase in their numbers is fraught with nutrient depletion in their habitat. For photosynthesizing cells, such regulation is unnecessary. It is not improbable that this is the cause of the fact that multicellular animals, retaining the ability to regulate the numbers of their cells, grow up to certain sizes, and their organs in regeneration (if any) are restored to their initial sizes. Conversely, plants, not having such regulation, grow virtually throughout their life and are incapable of regenerating in the generally accepted sense. One can assume that, in a culture of unicellular plants, there are no phase of a rapid decrease in the concentration and no difference between the concentrations of cells in the quasi-stationary phase in different volumes.

Explanations of five experimental facts marked in figures in the text of article are presented below.

According to our hypothesis, with growth of number of cells in volume the density of a stream of radiation grows. At achievement of that density of a stream which corresponds to concentration of cells to the beginning of quasi-stationary phase, the gradual removing of a part of cells from a cycle began, and in part of them most likely the apopthose mechanism is switched on [5.] (it is necessary to note, that in the literature we have not found the data on the phenomenon of apopthose in monocellulars. It is possible, that this method of destruction of "superfluous" cells as the mechanism of management of number of community has arisen at the earliest stages of evolution as well).

Approximately to the middle of the second day of development of culture there is a deviation from exponential growth that is shown in apparent increase of the average period of cell division. For some period of time two competing processes work: division of a part of cells which have not left yet a cycle, and a removing of cells from a cycle with compulsory destruction of a part from them. And the second process grasps the increasing number of cells. It results in approach of a phase of fast recession of concentration. When decreasing concentration of cells reaches that level to which the critical value of density of radiation corresponds, compulsory destruction of cells stops and culture moves to quasi-stationary phase. Reduction of concentration of cells in quasi-stationary phase most likely occurs due to natural destruction of cells, which average time of life is a number of days. Thus:

(1) fast recession of concentration of cells after achievement of a maximum of development of culture is caused not by an exhaustion of nutrients, but by switching on of the mechanism of regulation of number of cells;

(2) critical density of radiation in the big volumes is reached at smaller concentration of cells than in small volumes. Therefore in quasi-stationary phase concentration of cells in small volumes exceeds concentration of cells in the big volumes;

(3) according to A.G. Gurwitsch [1], radiation of dividing cells is capable to start division of other cells located close enough. Inside cultivation volume the density of this radiation, as shown previously, is proportional to the sizes of volume. Therefore, in the big volumes at early stages of development of culture concentration of cells grows faster, than in small;

(4) when we reduce cultivation volume in quasi-stationary phase, keeping in it the former concentration of cells, the density of a stream of radiation decreases, the block is switched off from cells and they start to divide until their concentration does not become sufficient for new switching on of the block;

(5) during the increase of cultivation volume in quasi-stationary phase with preservation of former concentration of cells, in a some period of time the mechanism destroying "superfluous" cells is switched on again.